\documentclass{osa-article}

\journal{oe}

\articletype{Research Article}

\begin{document}

\title{Coupling Light to Higher Order Transverse Modes of a Near-Concentric Optical Cavity}

\author{Adrian Nugraha Utama,\authormark{1} Chang Hoong Chow,\authormark{1}, Chi Huan Nguyen,\authormark{1} and Christian Kurtsiefer\authormark{1,2,*}}

\address{\authormark{1}Centre for Quantum Technologies, 3 Science Drive 2, Singapore 117543\\
\authormark{2}Department of Physics, National University of Singapore, 2 Science Drive 3, Singapore 117542}

\email{\authormark{*}phyck@nus.edu.sg} 



\begin{abstract}
Optical cavities in the near-concentric regime have near-degenerate transverse modes; the tight focusing transverse modes in this regime enable strong coupling with atoms. These features provide an interesting platform to explore multi-mode interaction between atoms and light. Here, we use a spatial light modulator (SLM) to shape the phase of an incoming light beam to match several Laguerre-Gaussian (LG) modes of a near-concentric optical cavity. We demonstrate coupling efficiency close to the theoretical prediction for single LG modes and well-defined combinations of them, limited mainly by imperfections in the cavity alignment.
\end{abstract}

\section{Introduction}

Transverse modes of paraxial beams are a set of unique field patterns perpendicular to the propagation of electromagnetic waves. They have a wide range of applications, such as increasing the information-carrying capacity in free-space~\cite{wang2012terabit} and fiber~\cite{bozinovic2013terabit, zhu2016encoding} communications, creating smaller focal volumes to achieve superresolution imaging~\cite{hasnaoui2011creation}, utilizing orbital angular momentum (OAM) for quantum key distribution~\cite{mafu2013higher}, and producing highly-entangled states~\cite{fickler2012quantum}. 
In optical cavities, transverse modes have been used to track atomic position via the observed mode pattern~\cite{horak2002optical, maunz2003emission, puppe2004single}, and to help enhancing the cooling process in atomic ensembles~\cite{gangl2000cooling, ritsch2013cold, black2003observation}. 
Optical cavities with near-degenerate transverse modes have also been used to engineer inter-mode coupling~\cite{klaassen2005transverse, benedikter2015transverse}, and to study crystallization domains in Bose-Einstein condensates (BEC)~\cite{gopalakrishnan2009emergent,  kollar2015adjustable, vaidya2018tunable, guo2019sign}.
Furthermore, transverse modes can be chosen as a degree of freedom for field quantization, along with wavelength and polarization, and can be utilized to explore atom-photon interaction as building blocks of a quantum network.

The near-degeneracy of transverse modes in an optical cavity arises in the region where the Gouy phase shifts of the cavity modes are fractions of $\pi$, notably in the confocal and concentric region~\cite{Saleh2001, papageorge2016coupling}. Cavity modes in the near-confocal region have relatively large mode volume, which is suitable to explore multi-mode interaction in large atomic ensemble such as BEC~\cite{kollar2015adjustable, papageorge2016coupling}. On the other hand, cavity modes in the near-concentric region have small mode volumes with a beam waist on the order of the atomic cross section, and thus show potential for strong interaction between light and single atoms~\cite{morin1994strong, durak2014diffraction, PhysRevA.96.031802}. 
The spatial resolution of the transverse modes can also be utilized to trap and couple selectively to small ensemble of single atoms. 
In centimetre-sized near-concentric cavities, the frequency spacing of the transverse modes ranges between $\sim 0.01$ to $1\,\mathrm{GHz}$ -- the lower limit is set by the last stable resonance from the critical point, which is less than half a wavelength away~\cite{PhysRevA.98.063833}. 
Unlike planar cavities, the frequency spacing is on the order of the hyperfine or the Zeeman level splitting of the atoms. This allows to explore single-quanta atomic nonlinearities with multiple optical modes coupled to different hyperfine or magnetic energy levels simultaneously, which has been previously demonstrated with two atomic transitions with $\sim 10\,\mathrm{THz}$ spacing using planar cavities~\cite{hamsen2018strong}. 

The transverse modes of a cavity can be excited by modifying the wavefront of the incoming Gaussian beam in a TE$_{00}$ mode to match the transverse spatial profile of the modes. 
In this work, we use a liquid-crystal spatial light modulator (SLM) to perform mode conversion by modulating the spatial phase profile. This enables coupling of a SLM-converted beam to a specific mode or a superposition of transverse modes in a near-concentric cavity. 
Furthermore, we examine how close to the critical point the transverse modes are still supported.  
Previously, such phase SLM have been utilized to excite the transverse modes of multimode fibers~\cite{dubois1994selective}, while excitation of cavity transverse modes in a near-confocal regime has been implemented with a digital micromirror device -- a type of binary-mask amplitude SLM~\cite{papageorge2016coupling}.
Compared with amplitude SLMs, phase SLMs can ideally perform mode conversion and coupling with higher overall efficiency as it does not require parts of the beam to be attenuated or diverted away. 
While the near-concentric cavities exhibit some technical complexities specific to the highly diverging modes in approaching the critical point, an efficient mode conversion enables interfacing of atomic qubits with multiple near-degenerate photonic modes. 

\section{Theory}

\subsection{Transverse modes of a cavity}
   
\begin{figure}[t]
\centering\includegraphics[width=12cm]{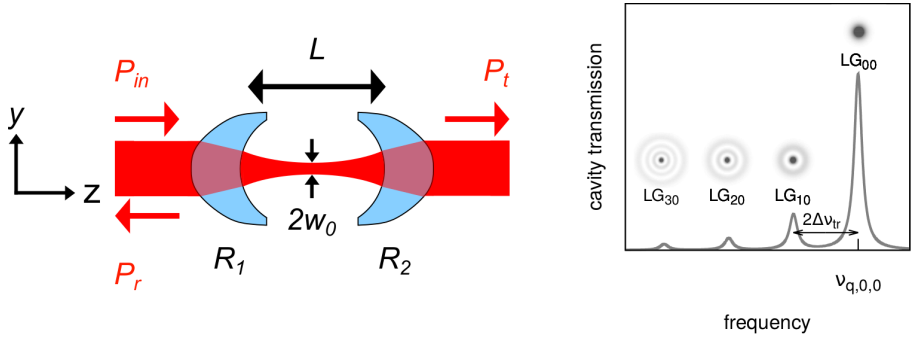}
\caption{Left: Schematic and coordinate system of the near-concentric cavity with a highly focused mode with $2 w_0$ waist diameter. Right: Example of the cavity transmission with input beam from a collimated fiber output. As the input beam has no orbital angular momentum ($l=0$), the frequency spacing between adjacent LG$_{\mathrm{m},0}$ modes would be $2 \Delta v_{tr}$. 
}
\label{fig:schematic}
\end{figure}

The spatial modes of the near-concentric cavity we investigate here are still well described by the paraxial approximation up to the last stable resonance~\cite{PhysRevA.98.063833}. 
We briefly present the theoretical framework to express paraxial transverse modes in an optical cavity with a scalar field that forms a standing wave~\cite{Saleh2001}. 
In a cylindrically symmetric cavity, the transverse mode profile can be described by a complex amplitude
\begin{equation}
U_{m,l}(\rho,\phi,z)= A_{l,m} \frac{w_0}{w(z)} \left(\frac{\rho}{w(z)}\right)^l \mathcal{L}^l_m \left(\frac{2\rho^2}{w^2(z)}\right) \exp \left(-\frac{\rho^2}{w^2(z)}\right) \exp\left(i \,\psi_{m,l}(\rho,\phi,z) \right) \ ,
\label{eq:lgexpr}
\end{equation}
where $m$ and $l$ are the radial and azimuthal mode numbers of the Laguerre-Gaussian (LG) beams, $A_{l,m}$ is the normalization constant, $w(z)=w_0\sqrt{1+(z/z_0)^2}$ is the beam radius along the $z$ direction with $z_0=\pi w_0^2/\lambda$ as the Rayleigh range and $w_0$ as the waist radius, $\mathcal{L}^l_m$ is the generalized Laguerre polynomial, and $\psi_{m,l}(\rho,\phi,z)$ is the real-valued phase of the LG beam, given by
\begin{equation}
\psi_{m,l}(\rho,\phi,z) = -kz -k\frac{\rho^2}{2R(z)} -l\phi + (2m+l+1)\zeta(z) \ ,
\label{eq:phaseLG}
\end{equation}
where $R(z)=z+z_0^2/z$ is the curvature radius of the wavefront, and $\zeta(z)=\tan^{-1}(z/z_0)$ is the Gouy phase.

Inside a cavity, the LG modes are bounded by the two spherical mirror surfaces of radii $R_1$ and $R_2$ spaced $L$ apart. The modes are geometrically stable when stability parameters $g_1 = 1 - L/R_1$ and $g_2 = 1 - L/R_2$ satisfy the confinement condition $0\leq g_1 g_2 \leq 1$~\cite{fox1963modes}. In symmetric cavities ($g_1=g_2=g$), the marginally stable concentric mode is obtained for a critical mirror separation of $L=2R$ and $g=-1$. 
Near-concentric modes depart from this point towards the stable region -- the distance away from the critical mirror separation is characterized by the critical distance $d=2R-L$, with $g=-1+d/R$.

The resonance frequencies of the cavity depend on the transverse mode numbers $m$ and $l$,
\begin{equation}
v_{q,m,l} = \left(q + (2m+l+1)\frac{\Delta\zeta}{\pi}\right) v_F  \ ,
\end{equation}
where $q$ is the longitudinal mode number of the cavity, $v_F=c/2L$ is the cavity free spectral range, and $\Delta\zeta=\zeta(z_{M2})-\zeta(z_{M1})$ is the Gouy phase difference between the two cavity mirrors. In near-concentric symmetrical cavities, the frequency spacing between two consecutive transverse modes is given by 
\begin{equation}
\Delta v_{tr} =  v_{q,0,0} - v_{q-1,0,1} = \frac{v_F}{\pi}  \cos^{-1} \left( 1- \frac{d}{R}\right) \ ,
\end{equation}
where $\Delta v_{tr}\rightarrow 0$ as $d\rightarrow 0$. By measuring the frequency separation between the transverse modes, we can estimate the critical distance $d$ and deduce the waist radius $w_0$~\cite{PhysRevA.98.063833}. 

\subsection{Atom-light coupling in near-concentric cavity}

The strength of atom-light interaction is characterized by the coupling constant $g_{ac}\propto d_a/ \sqrt{V_m}$, which depends on the atomic dipole moment $d_a$ and the effective mode volume $V_m \approx \pi w_0^2 L$~\cite{aljunid2011interaction}. 
Small mode volumes can be achieved either with short cavity length $L$ or small waist radius $w_0$. 
Due to small $w_0$ in approaching the critical point, near-concentric cavities exhibit strong atom-light coupling strength $g_{ac}$ comparable to $\mathrm{\mu m}$-length cavities or fiber cavities~\cite{PhysRevA.96.031802}.
  
In addition, all the radial transverse modes (LG modes with $l=0$) at a particular critical distance $d$ have identical effective mode volumes, as the normalization relation $\int_0^{\infty} e^{-u} \left[ \mathbb{L}^0_m (u) \right]^2 du = 1$ yields the same prefactor $A_{0,m}$ in Eq.~\ref{eq:lgexpr}. This allows coupling between an atom and cavity modes with equal strength across all radial transverse modes. 

\subsection{Mode matching to a cavity}
\label{section:modeMatching}

We briefly describe the method to measure the mode matching efficiency in a cavity with realistic losses, following the cavity characterization technique in Ref~\cite{PhysRevA.64.033804}. 
The power transmission through a cavity with mirrors of the same reflectivity is given by
\begin{equation}
T(\omega) =\frac{P_{t}(\omega)}{P_{in}}= \eta \frac{\kappa_m^2}{(\kappa_m+\kappa_l)^2+(\omega-\omega_0)^2} \ ,
\label{eq:transSpec}
\end{equation}
where $P_{t}(\omega)$ is the light power transmitted through the cavity, $P_{in}$ is the input power, $\eta$ is the spatial mode matching efficiency, $\omega_0$ is the cavity resonance frequency, and $\kappa_m$ and $\kappa_l$ are the cavity decay rates due to the mirror transmission and scattering losses, respectively. On the other hand, the fraction of power reflected back from the cavity is given by
\begin{equation}
R(\omega) =\frac{P_{r}(\omega)}{P_{in}}= 1 - \eta \frac{\kappa_m^2+2\kappa_m\kappa_l}{(\kappa_m+\kappa_l)^2+(\omega-\omega_0)^2} \ ,
\label{eq:refSpec}
\end{equation}
where $P_{r}(\omega)$ is the light power reflected by the cavity. The total cavity decay rate, $\kappa = \kappa_m + \kappa_l$, determines the cavity finesse, $F= \pi /\kappa v_F $, and can be obtained by fitting Eq.~(\ref{eq:transSpec}) to the measured transmission spectrum. 

The mode matching efficiency $\eta$ can be obtained from Eq.~(\ref{eq:transSpec}) and Eq.~(\ref{eq:refSpec}) on the cavity resonance ($\omega=\omega_0$),
\begin{equation}
\eta =  \frac{(1+\alpha)^2}{(2\alpha)^2}T(\omega_0) \ ,
\label{eq:findEta}
\end{equation}
where $\alpha = \kappa_m/(2\kappa_l +\kappa_m)$ is determined by the cavity decay rates, and thus is a physical property of the cavity mirrors -- for mirrors with no scattering or absorption losses, $\alpha=1$. The parameter $\alpha$ can be estimated from the measurement of the  cavity transmission and reflection at resonance:
\begin{equation}
\alpha =  \frac{T(\omega_0)}{1-R(\omega_0)} \ ,
\label{eq:findAlpha}
\end{equation}
which represents the effectiveness of the cavity transmission. The cavity decay rates can be obtained as $\kappa_m= 2\kappa \alpha/(1+\alpha)$ and $\kappa_l= \kappa (1-\alpha)/(1+\alpha)$ from measured values of $\kappa$ and $\alpha$.

\subsection{Beam shaping with SLM}
\label{section:beamShaping}

To prepare LG beams and couple to the transverse modes of the near-concentric cavity, we use a liquid-crystal phase SLM to perform mode conversion from a collimated single mode fiber ouput (approximating a Gaussian beam).
Such a transformation can be performed with a spatial filter which modulates both the amplitude and the phase of the incoming mode, and described by a generalized filter function $T(\mathbf{x}) = M(\mathbf{x}) \exp(i \,\Phi(\mathbf{x}))$. However, a liquid-crystal SLM only modulates the phase of the incoming beam and hence only provides the transformation $T(\mathbf{x}) = \exp(i \,\Phi(\mathbf{x}))$. 

There are several methods to perform both amplitude and phase modulation using only a phase SLM. In one method, the SLM can be operated in a phase-grating configuration -- this produces both the carrier and first-order diffraction beams, where phase and amplitude can be varied using the modulation angle and the modulation depth, respectively~\cite{kirk1971phase, davis1999encoding}. This method typically requires a high-resolution SLM to encode the phase and amplitude information sufficiently precise with the phase grating. However, recent works explored encoding techniques with different sets of amplitude modulation bases 
which allow the usage of a low-resolution phase SLM~\cite{arrizon2007pixelated, ando2009mode, clark2016comparison, forbes2016creation}. 
Another method relies on using two SLMs with a polarizer to modulate the amplitude and phase of the incoming beam independently~\cite{juday1991full, neto1996full, reichelt2012full}.

Here, we use a much simpler technique that does not require parts of the beam to be diverted away or attenuated, because LG modes with relatively high purity can be created by spatially modulating the incoming Gaussian beam with only the phase component of the desired LG modes~\cite{arlt1998production, ohtake2007universal, matsumoto2008generation}. The cavity then acts as a filter to attenuate the remaining off-resonant LG mode components, while transmitting the desired LG mode. The SLM phase function for this transformation is given by
\begin{equation}
 \Phi(\mathbf{\rho,\phi}) = \arg \left[ U_{m,l}(\rho,\phi,0) \right]= \arg \left[ \mathcal{L}^l_m \left(\frac{2\rho^2}{w^2}\right) \right] - l \phi \ ,
\label{eq:SLMphasefunction}
\end{equation}
with the incoming Gaussian mode $U_0(\rho) = A_0 \exp \left(-\rho^2/w_0^2\right)$. The ratio $w/w_0$ can be varied to optimize the mode overlap of the resulting SLM output mode to the particular LG mode -- the mode overlap is defined as $\int (d\sigma) U_1^*(\rho,\phi) U_2(\rho,\phi)$, and the modulus square of the mode overlap is equivalent to the mode matching efficiency $\eta$ as defined in Section~\ref{section:modeMatching}. 
For relatively small mode indices $m$ and $l$, the mode matching efficiencies of the same LG modes are relatively high, with low mode matching efficiencies to different LG modes (see Table~\ref{table:mode_overlap}). 
Due to the simplicity of the phase function, this technique can also be implemented using physical phase plates~\cite{sueda2004laguerre,bencheikh2014generation}.

\begin{table}[t]
\centering
 \begin{tabular}{|c |c |c c c c c c|} 
 \cline{3-8}
  \multicolumn{2}{c|}{} &\multicolumn{6}{c|}{Mode matching efficiencies} \\ 
\hline
 SLM output & $W/W_0$ & LG$_{00}$ & LG$_{10}$ & LG$_{20}$ & LG$_{30}$ & LG$_{40}$ & LG$_{50}$\\ 
 \hline
 LG$_{10}$ & 0.57 & 0.1\% & \textbf{81.2\%} & 0.0\% & 2.4\% & 1.3\% & 0.7\%\\ 
 LG$_{20}$ & 0.45 & 1.3\% & 0.1\% & \textbf{76.9\%} & 0.1\% & 1.6\% & 4.5\%\\ 
 LG$_{30}$ & 0.39 & 0.4\% & 1.2\% & 0.5\% & \textbf{74.6\%} & 0.3\% & 0.9\%\\ 
 LG$_{40}$ & 0.35 & 0.2\% & 0.4\% & 1.2\% & 0.8\% & \textbf{73.2\%} & 0.5\%\\ 
 \hline
 \end{tabular}
 \caption{Calculated values of the mode matching efficiencies $\eta$ between the SLM output and the LG modes for $l=0$, up to LG$_{50}$ -- mode matching to modes higher than LG$_{50}$ are smaller (not shown in table), and the cumulative efficiencies sum up to unity asymptotically. The model does not incorporate pixellation and aperture effects caused by a real SLM. 
  }
 \label{table:mode_overlap}
\end{table}

\section{Experiment}

\subsection{Experimental setup}

\begin{figure}[t]
\centering\includegraphics[width=12cm]{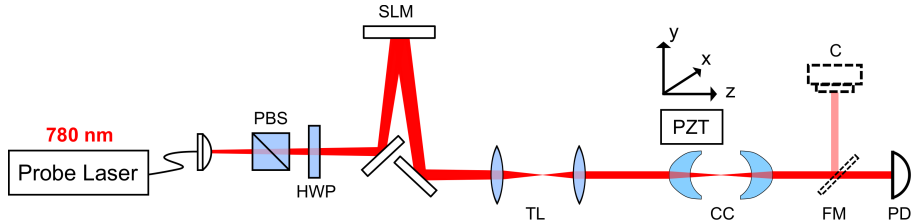}
\caption{Optical setup. A spatial light modulator (SLM) transforms light emerging from a single mode optical fiber to match the LG modes of the near concentric cavity (CC). 
A telescope (TL) facilitates mode matching between the SLM output and the cavity. Cavity transmission is monitored using either a photodetector (PD) or mode camera (C), selected by a flip mirror (FM).}
\label{fig:setup}
\end{figure}

The design and construction of the near-concentric cavity was described previously~\cite{durak2014diffraction, PhysRevA.98.063833}. The anaclastic lens-mirror design allows highly divergent modes of the near-concentric cavity to be transformed into collimated modes with a single element. This simplifies the requirement of the optical components to generate and measure collimated LG beams on the input and output of the cavity (see Figure~\ref{fig:setup}). 

\subsubsection{Mode conversion with SLM}
\label{sec:modegen}

We use a liquid-crystal SLM (Meadowlark HV 512 DVI) with an active area of 12.8 mm $\times$ 12.8 mm and resolution of 512x512 pixels.
As this SLM only modulates light with a particular linear polarization, a sequence of a polarizing beam-splitter (PBS) and a half-wave plate (HWP) prepares the correct polarization to match the SLM polarization axis. 
We minimize the pixelation artifact by using a significant portion of the SLM area. To achieve this, we prepare a slightly divergent beam with beam diameter ($1/e^2$ width) ranging from 3 to 7 mm, measured at the SLM.

The phase modulation applied on the SLM consists of three components: the LG mode-generating phase pattern as described in Eq.~\ref{eq:SLMphasefunction}, the correction phase pattern provided by the manufacturer, and a quadratic phase pattern which effectively acts as a Fresnel lens with variable focal length. This SLM-generated Fresnel lens helps in supressing the unmodulated light on the SLM output (more commonly done with a blazed grating pattern~\cite{matsumoto2008generation}).
In addition, the combination of the Fresnel lens with a telescope of variable length and magnification creates a collimated LG beam with tunable beam size. The appropriate values for the Fresnel lens and telescope parameters are obtained with ray-tracing simulations.

\subsubsection{Cavity alignment}
In the cavity design~\cite{PhysRevA.98.063833}, one cavity mirror is placed on 3D piezo translation stage (Figure~\ref{fig:setup}) to allow for both the longitudinal (z direction) and transverse alignment (x and y directions). The longitudinal alignment changes the cavity length to be resonant to a particular optical frequency, while the transverse alignment is performed to establish cylindrical symmetry of the system. 
Small tip-tilt misalignment can also be corrected by the transverse alignment, if the mirrors are perfectly spherical. However, such a correction misaligns the two anaclastic lens-mirror axes from the cavity axis, resulting in slightly asymmetric collimated output modes.

The transmission and reflection spectrum of the cavity are obtained by measuring the light intensity with a photodetector while varying the cavity length linearly over time. The detuning from the cavity resonance is expressed is corresponding units of light frequency -- the conversion factor is determined by measuring the spacing of the frequency sideband generated with an electro-optical modulator. 

\subsubsection{Measurement of the mode matching efficiency}
\label{sec:modematch}

The mode matching efficiency $\eta$ (Eq.~\ref{eq:findEta}) quantifies how well the input mode couples to the cavity mode. It only depends on the resonant power transmission at resonance $T(\omega_0)$ and the effective transmission coefficient $\alpha$ (Eq.~\ref{eq:findAlpha}). We characterize the value of $\alpha$ by coupling a Gaussian beam (from a collimated single mode fiber output mode) into the cavity without the SLM. The transmission and reflection spectrum were recorded. From the fitting, we obtain $T(\omega_0)=19.5(1)\%$, $R(\omega_0)=33.6(2)\%$, and $\kappa= 2\pi \times 24.8(8)\,\mathrm{MHz}$. From these parameters, we estimate $\alpha=0.294(2)$, which results in a mode matching efficiency of $\eta=94(1)\%$ for Gaussian beam, and cavity decay rates of $\kappa_m= 2\pi \times 11.3(4)\,\mathrm{MHz}$ and $\kappa_l= 2\pi \times 13.5(4)\,\mathrm{MHz}$.

To estimate the mode matching efficiencies for SLM-generated LG modes, we obtain the cavity transmission spectrum $T(\omega)$ and multiply it with $(1+\alpha)^2/(2\alpha)^2$ (the prefactor in Eq.~\ref{eq:findEta}) to obtain the mode transmission spectrum $\eta(\omega)$. We fit this spectrum with a Lorentzian profile, and estimate the mode matching efficiency $\eta=\eta(\omega_0)$ from the fit amplitude.
The parameters from the ray-tracing simulation helps to start the coupling procedure, and we fine-tune these values further to maximize the mode matching efficiency. 

\subsection{Mode-matching to single LG modes}
\label{sec:singlemode}

\begin{figure}[t]
\centering
\includegraphics[width=6.5cm]{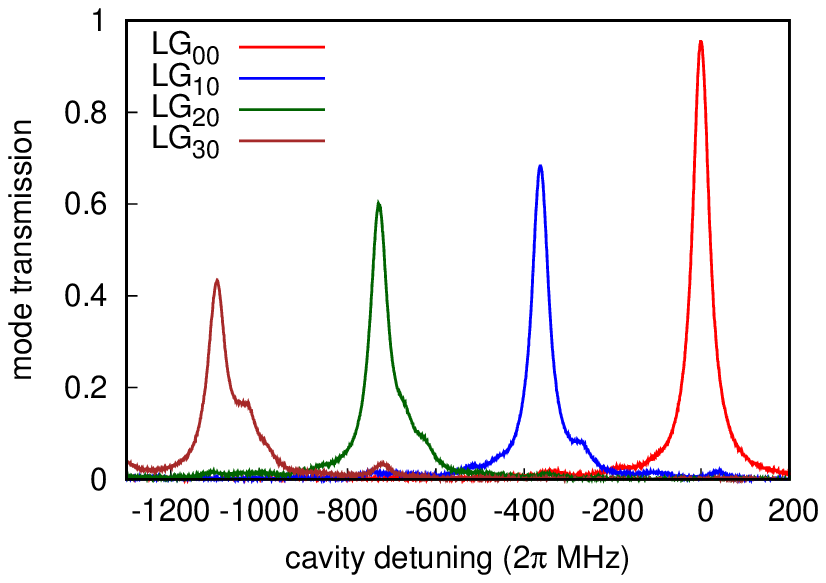}
\includegraphics[width=4.4cm]{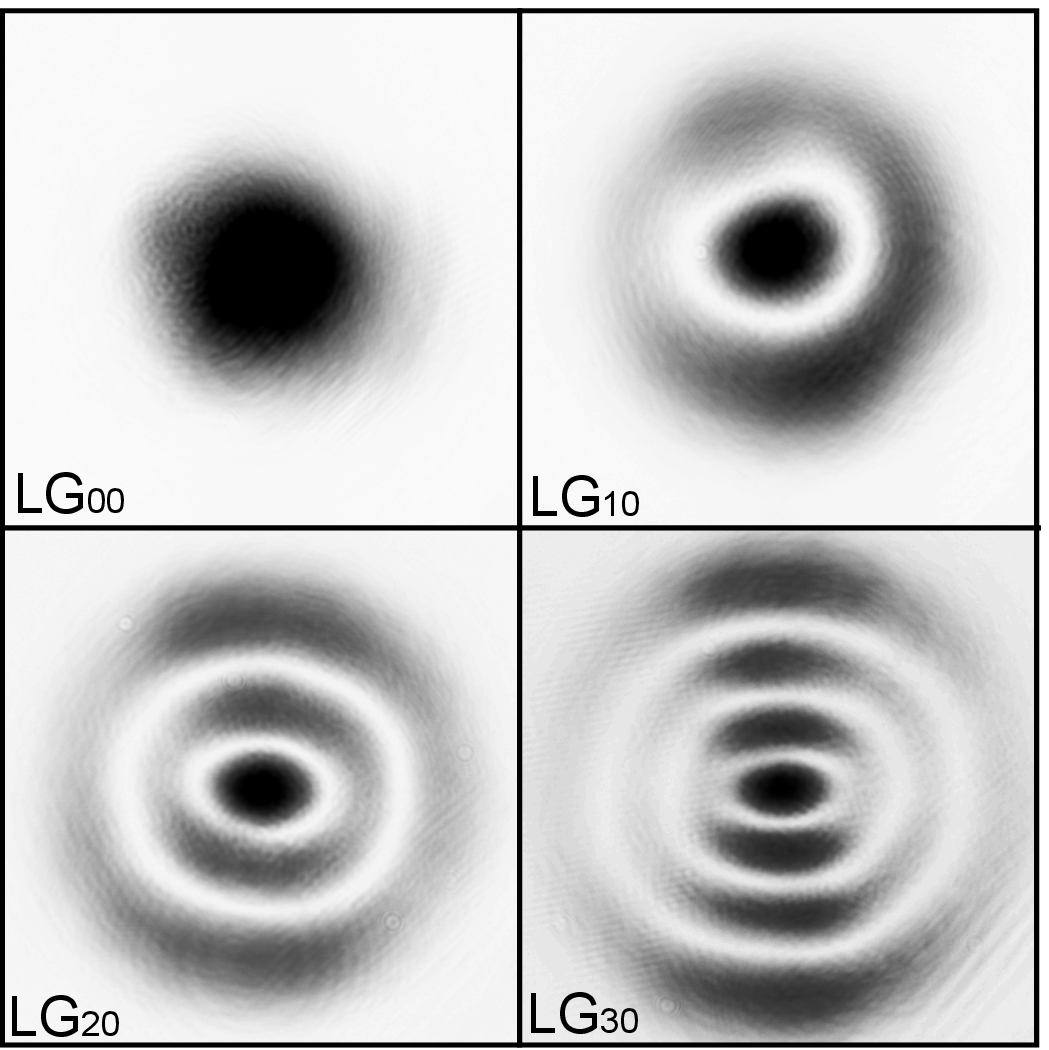}
\caption{Left: measured cavity transmission for radial LG modes ($l=0$). The detuning is defined with respect to the LG$_{00}$ resonance; higher order the modes are spaced $2\Delta v_{tr}$ apart. Right: the corresponding cavity output mode observed with the mode camera.}
\label{fig:single_l0}
\end{figure}

\begin{figure}[t]
\centering
\includegraphics[width=6.5cm]{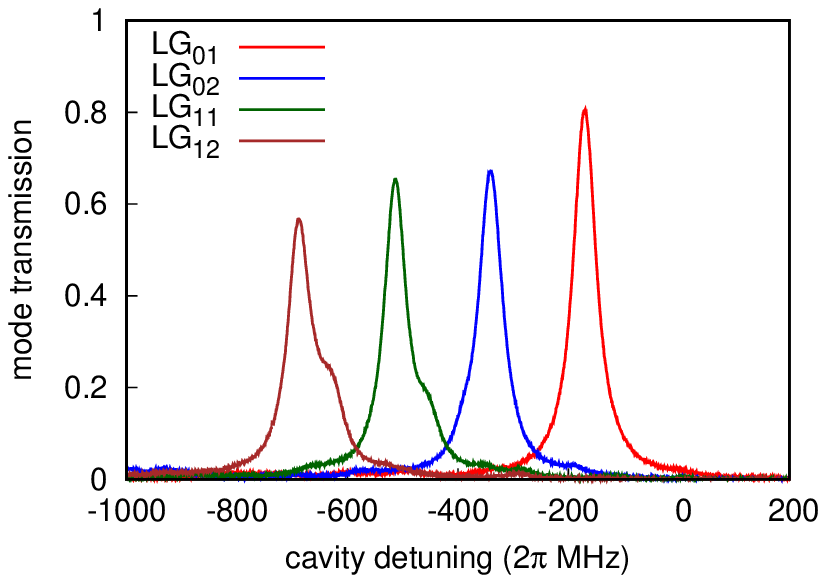}
\includegraphics[width=4.4cm]{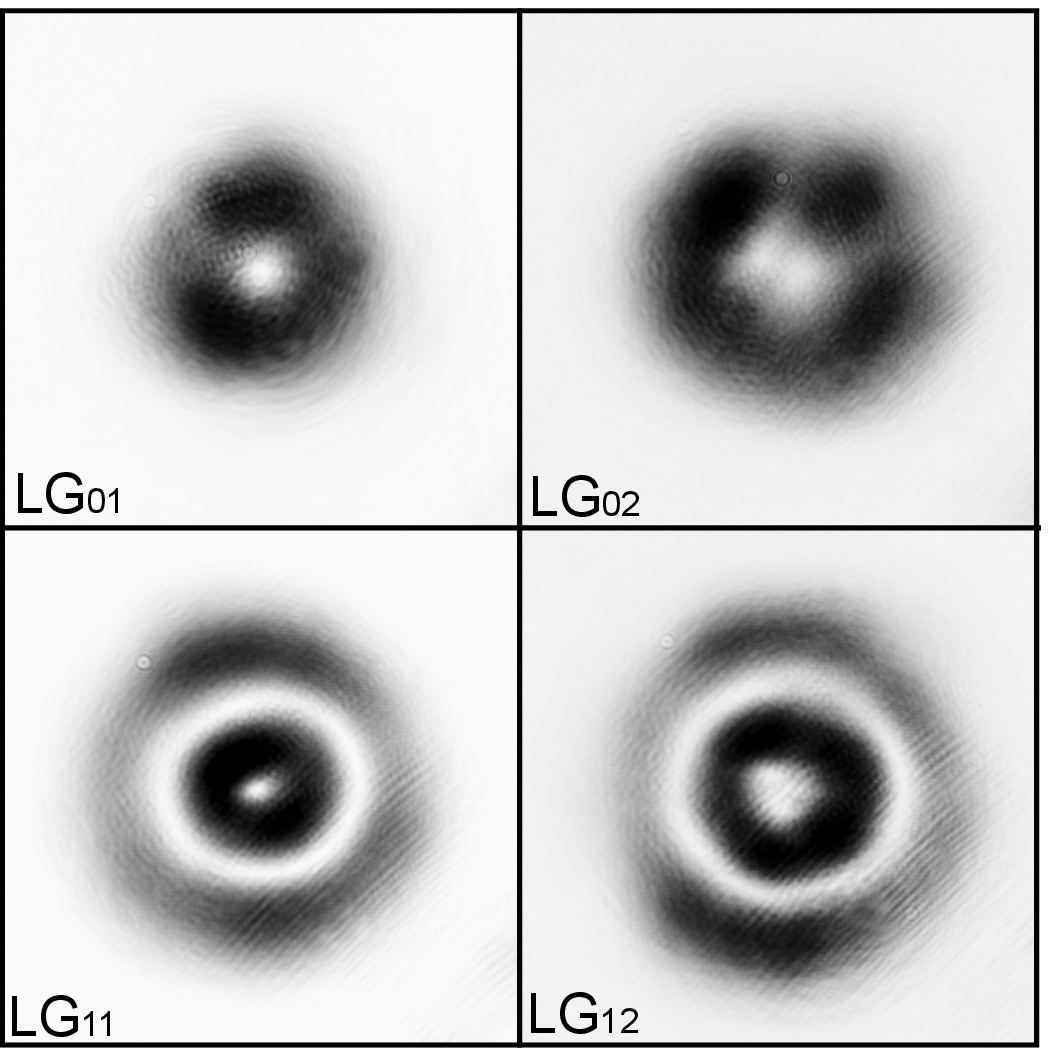}
\caption{Left: measured cavity transmission for LG modes with low angular momentum ($l=1$ and $l=2$). Right: the corresponding cavity output mode.}
\label{fig:single_ln0}
\end{figure}

\begin{table}[t]
\centering
 \begin{tabular}{|c c c |} 
\hline  
 Mode & Sim. & Exp.\\ 
 \hline
 LG$_{00}$ & 100\% & 96(1)\%  \\ 
 LG$_{10}$ & 81.2\% & 68(1)\%  \\ 
 LG$_{20}$ & 76.9\% & 57(1)\%  \\ 
 LG$_{30}$ & 74.7\% & 38(1)\%  \\ 
 \hline
 \end{tabular}
\quad
 \begin{tabular}{|c c c |} 
\hline
 Mode & Sim. & Exp. \\ 
 \hline
 LG$_{01}$ & 93.1\% & 81(1)\%  \\ 
 LG$_{02}$ & 84.4\% & 67(1)\%  \\ 
 LG$_{11}$ & 81.8\% & 63(1)\%  \\ 
 LG$_{12}$ & 79.8\% & 53(1)\%  \\ 
 \hline
 \end{tabular}
 \caption{Comparison of mode matching efficiencies between simulation and experiment for single LG modes.}
 \label{table:mode_eff_sm}
\end{table}

We generate a single LG mode using the SLM and couple it to the near concentric cavity. 
The cavity is located at a critical distance of $d = 4.8(2) \mathrm{\mu m}$ with $g = -0.99912(4)$, corresponding to a measured transverse mode spacing of $\Delta v_{tr} = v_F  (1-\Delta\zeta / \pi) = 182(5)\,\mathrm{MHz}$ between adjacent LG modes. 
The cavity spectra and the camera-captured output modes are depicted in Figure~\ref{fig:single_l0} for LG modes with no angular momentum ($l=0$), and in Figure~\ref{fig:single_ln0} for LG modes with angular momentum ($l\neq 0$). The measured mode matching efficiencies are close to the simulated values (see Table~\ref{table:mode_eff_sm}), although they decrease with higher mode numbers. We attribute this to limited SLM pixel resolution, axial mismatch between the cavity and the anaclastic lens axis due to tip-tilt misalignment, and a mirror surface deviation from a perfect spherical profile. These factors also contribute to some irregularities on the output mode observed by the mode camera.

\subsection{Mode-matching to a superposition of LG modes}

Superpositions of transverse modes in a cavity provide an interesting avenue to explore multi-photon interaction with atomic medium~\cite{hamsen2018strong}.
We demonstrate the coupling of the SLM-generated beam to an arbitrary superposition of LG modes. We use the method described in Section~\ref{sec:modegen} by considering the resultant mode as a superposition of individual LG modes,
\begin{equation}
U_{res} = \sum A_{l,m} \exp\left(i\phi_{l,m}\right) LG_{lm} \ , 
\end{equation}
where $A_{l,m}$ is the amplitude of each constituting LG mode and $\phi_{l,m}$ is the relative phase of the LG mode. 

\begin{figure}[t]
\centering
\includegraphics[width=6.5cm]{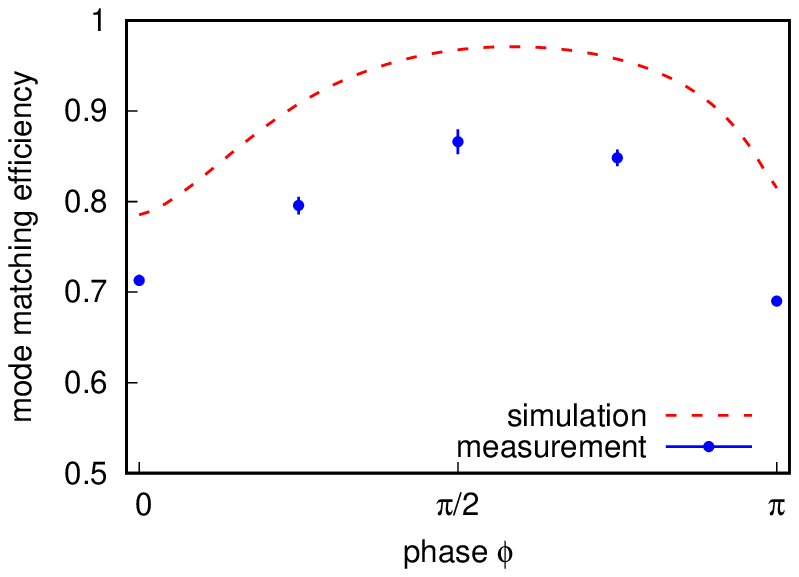}
\includegraphics[width=6.5cm]{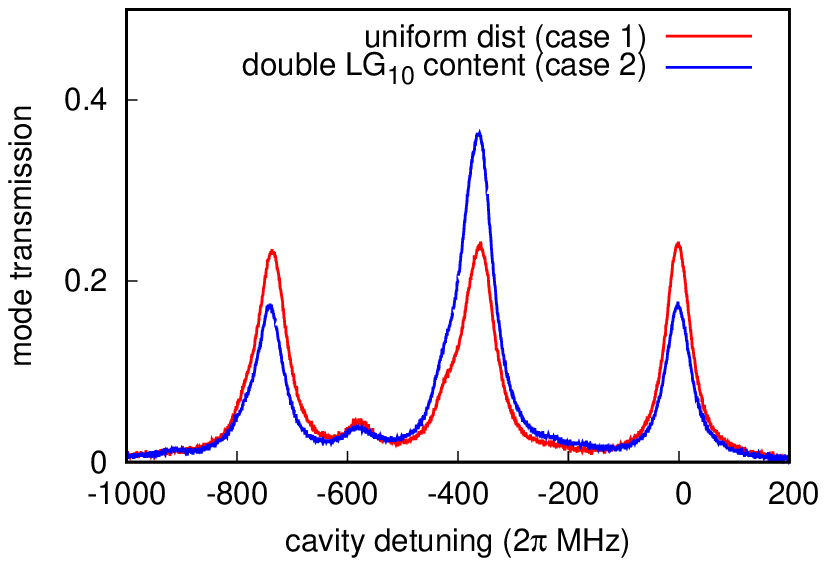}
\caption{Left: Coupling to equal parts of LG$_{00}$ and LG$_{10}$ while varying their phase difference. 
Right: Coupling to a superposition of LG$_{00}$, LG$_{10}$ and LG$_{20}$.}
\label{fig:multimode}
\end{figure}

Figure~\ref{fig:multimode} (left) shows the mode matching efficiency in coupling the SLM-generated beam to the cavity superposition mode $U_{\{00,10\}}=\left(LG_{00} + e^{i\phi} LG_{10} \right)/\sqrt{2}$ with a varying relative phase angle $\phi$. To obtain a balanced distribution of LG$_{00}$ and LG$_{10}$, we introduce a mode amplitude $A_{10}$ to the SLM spatial phase pattern,
\begin{equation}
\Phi = \arg \left[ U_{\{00,10\}} \right] = \arg \left[ \frac{LG_{00} + A_{10} e^{i\phi} LG_{10} }{\sqrt{1+A_{10}^2}} \right] \ , 
\end{equation} 
and vary the amplitude $A_{10}$ and $w/w_0$, maximising the mode matching efficiency subject to the balanced distribution constraint. The mode matching efficiency $\eta$ is obtained by adding the mode transmission amplitudes of both the $LG_{00}$ and $LG_{10}$ modes, while ensuring that they are balanced within $\sim1\%$. 
The measured values follow a similar trend with the simulated values, with some offset ($\sim 10\%$) attributable to the SLM pixel size and the mirror irregularities as described previously.
The highest mode matching efficiency ($\eta=87(1)\%$) occurs around $\phi=\pi/2$, in which case the in-phase component of the beam encodes the $LG_{00}$ mode, while the quadrature component of the beam encodes the $LG_{10}$ mode.

Figure~\ref{fig:multimode} (right) shows the transmission spectra of a superposition of three modes. Modes LG$_{00}$, LG$_{10}$, and LG$_{20}$ are superposed with a relative phase difference of $2\pi/3$ to distribute the phases evenly on the complex plane. The corresponding SLM spatial pattern is given by
\begin{equation}
\Phi = \arg \left[ U_{\{00,10,20\}} \right] = \arg \left[ \frac{LG_{00} + A_{10} e^{i2\pi/3} LG_{10} + A_{20} e^{i4\pi/3} LG_{20} }{\sqrt{1+A_{10}^2+A_{20}^2}} \right] \ ,
\end{equation} 
where $A_{10}$, $A_{20}$ and $w/w_0$ are parameters to be varied to obtain the desired mode distribution and the efficiency. Two cases are illustrated in Figure~\ref{fig:multimode} (right): (1) equally distributed modes, i.e. $U_{\{00,10,20\}}=\left(LG_{00} + e^{i2\pi/3} LG_{10} + e^{i4\pi/3} LG_{20}\right)/\sqrt{3}$, and (2) LG$_{10}$ content double the content of the other modes, i.e. $U_{\{00,10,20\}}=\left(LG_{00} + \sqrt{2}e^{i2\pi/3} LG_{10} + e^{i4\pi/3} LG_{20}\right)/2$. The theoretical efficiencies under optimized parameters are $95.6\%$ and $97.2\%$ for case (1) and (2), while the measured efficiencies are $71(1)\%$ and $70(1)\%$, respectively.
We attribute this discrepancy to the imperfections of the SLM and cavity as described previously, and in particular when coupling to the superposition component with higher mode numbers.

\subsection{Mode-matching at different critical distances}

Small critical distances provide strong field focusing and a small mode volume. In addition, the frequency spacing of the transverse modes decreases with smaller critical distances, leading to the mode degeneracy at the critical point~\cite{PhysRevA.98.063833}. We study how the mode matching of a single LG mode performs at different critical distances. We use the SLM to couple to LG$_{00}$, LG$_{10}$, and LG$_{20}$ modes of the cavity, and obtain the cavity transmission spectra. We find that the linewidth of the cavity spectra increases for smaller critical distances, while the mode transmission amplitude decreases. This is likely due to diffraction losses as the cavity approaches the critical point. 

\begin{figure}[t]
\centering
\includegraphics[width=6.5cm]{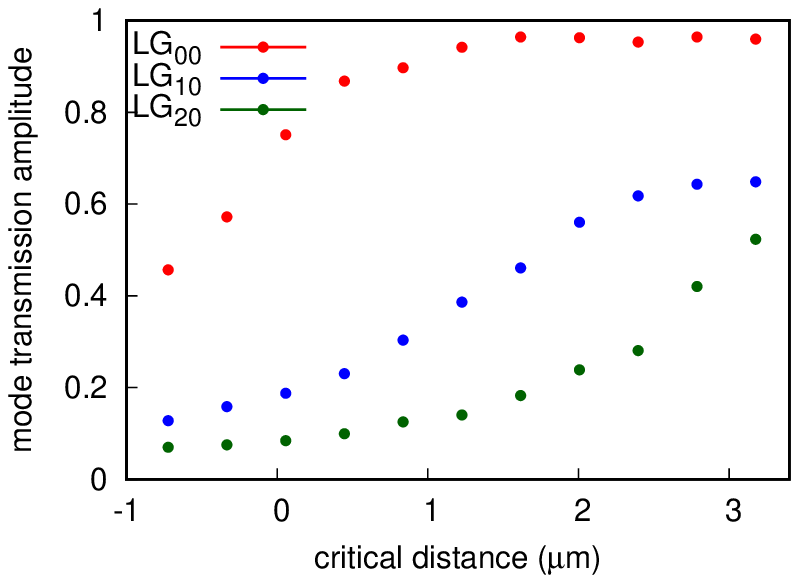}
\includegraphics[width=6.5cm]{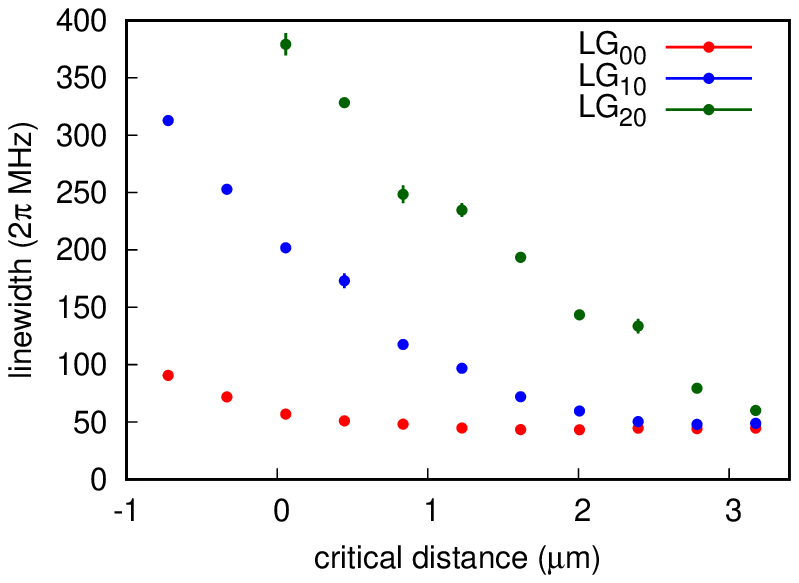}
\caption{Left: transmission amplitude of different LG modes over a range of critical distances. Right: The corresponding linewidth (FWHM).}
\label{fig:criticaldist}
\end{figure}

\begin{figure}[t]
\centering
\includegraphics[width=6cm]{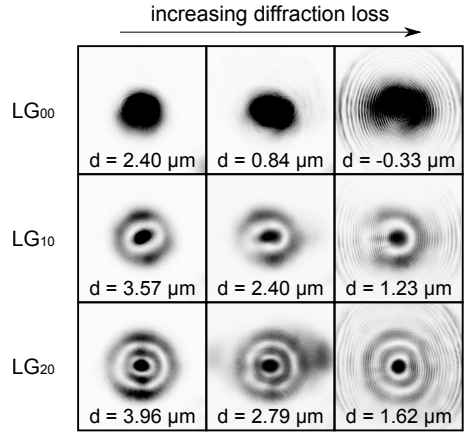}
\caption{The cavity modes observed with the mode camera.
For a small critical distance, diffraction loss becomes significant and distorts the mode profile.
The diffraction rings are caused by the aperture of the anaclastic lens. }
\label{fig:criticalcam}
\end{figure}

The critical distance can be estimated from the transverse mode spacing. By changing the cavity length and keeping the laser frequency fixed, we obtain neighbouring cavity spectra spaced $\Delta d=\lambda/2$ apart.
Figure~\ref{fig:criticaldist} shows the cavity transmission amplitudes and the cavity linewidths for various critical distances.
Without diffraction loss, the mode transmission amplitude is equivalent to the mode-matching efficiency $\eta$. 
However, as the diffraction loss increases, the effective transmission coefficient $\alpha$ also changes. Hence, the mode transmission amplitude describes the mode-matching efficiency weighted by a factor associated with the diffraction loss. 
In the high diffraction loss regime, it becomes hard to couple to a particular lossy eigenmode, and characterize its linewidth to obtain $\alpha$, as different transverse modes start to overlap in frequency.
Figure~\ref{fig:criticalcam} shows the spatial profile of the cavity transmission, captured with the mode camera. Diffraction rings become visible at the critical distance where the linewidth increases.

The near-concentric cavity can support several LG modes reasonably close ($\sim$ a few $\mu$m) to the critical point. However, higher order LG modes start to exhibit diffraction losses at larger critical distances, due to larger LG beam sizes.
The performance of the cavity mirrors can be characterized with an effective aperture -- for every round trip, the cavity mode is clipped by a circular aperture with diameter $a$ on the mirror, effectively blocking some outer parts of the beam. As a first order approximation, we assume the LG modes to be unperturbed after subsequent round trips. 
To estimate the onset of the diffraction loss, we choose an aperture size to block $\sim 1\%$ of the mode (the  diffraction loss is $2\kappa_{ap} \sim 2\pi \times 20\,\mathrm{MHz}$), which on the same order as the mirror transmission and scattering losses. 
From Figure~\ref{fig:criticaldist} (right), the effective aperture diameter is estimated to be $a_{\mathrm{exp}} = 1.40(6) \,\mathrm{mm}$ with the onset of the diffraction loss at critical distances of $0.46(8) \,\mathrm{\mu m}$ for LG$_{00}$, $1.8(3) \,\mathrm{\mu m}$ for LG$_{10}$, and $3.8(6) \,\mathrm{\mu m}$ for LG$_{20}$. 

The estimated effective aperture $a_{\mathrm{exp}} = 1.40(6) \,\mathrm{mm}$ is comparatively lower than the nominal aperture of the anaclastic lens-mirror design $a_{\mathrm{nom}} = 4.07 \,\mathrm{mm}$.
We suspect this to be due to a combination of:  
(1) local aberrations of the mirror surface due to mechanical stresses induced by the temperature change and the clamping process~\cite{legero2010tuning,yoder2008mounting}, (2) angle-dependent variation on the wavefront due to the multi-layered coating~\cite{kleckner2010diffraction}, and (3) the validity of the paraxial approximation for strongly diverging modes~\cite{chen2002analyses}, particularly for higher orders.
By slightly modifying the mirror shape or the coating layers, it might be possible to increase the effective aperture of the cavity and obtain stable LG modes even closer to the critical point.

\section{Conclusion}
In summary, we presented a mode-matching procedure to excite several transverse modes of a near-concentric cavity with a relatively high conversion efficiency. We use an SLM to engineer the spatial phase of an input Gaussian beam to selectively match a specific LG mode, and observe experimental
mode matching efficiencies close to theoretical predictions for several low-order LG modes, despite the imperfections in the cavity alignment and mirror surface, and the limited resolution of the SLM.
We demonstrated that a superposition of cavity modes can be generated with a high fidelity, and showed that a near-concentric cavity can support several LG modes up to critical distances of a few $\mathrm{\mu m}$ before the diffraction loss dominates.

The near-concentric regime of an optical cavity supports transverse modes which are spaced close to one another, on the same order of the magnetic level or hyperfine splitting of the atoms. 
Exciting the transverse modes in such a regime is a step towards exploring interaction between atoms and strongly focused near-degenerate spatial modes.
The nonlinearity arising from multiple photons interacting with single atoms can therefore provide a building block for scalable quantum networks.

\section*{Funding}
This research is supported by the Research Centres of Excellence programme supported by the National Research Foundation (NRF) Singapore and the Ministry of Education, Singapore.


\end{document}